\documentclass[aps,twocolumn,showpacs,floatfix,prd]{revtex4-1}

\usepackage{bm,enumerate,dcolumn,graphicx,color,amsmath,amssymb}

\begin{document}

\title{
Dark forces and atomic electric dipole moments
}
\author{Heman Gharibnejad}
\affiliation{Department of Physics, University of Nevada, Reno, NV 89557, USA}
\author{Andrei Derevianko}
\affiliation{Department of Physics, University of Nevada, Reno, NV 89557, USA}

\date{\today}

\begin{abstract}
Postulating the existence of a  finite-mass mediator of T,P-odd coupling between atomic electrons and nucleons we consider its effect on
permanent electric dipole moment (EDM) of diamagnetic atoms. We present both numerical and analytical analysis for such mediator-induced EDMs and compare it with EDM results for the conventional contact interaction. Based on this analysis we
derive limits on coupling strengths and  carrier masses from experimental limits on EDM of $^{199}$Hg atom.
\end{abstract}

\maketitle

\section{Introduction}
\label{Sec:Intro}

The observational evidence for dark matter indicates the intriguing possibility of a ``dark sector'' extension to the Standard Model (SM). Dark matter in fact may be a small part of the dark sector or indeed many dark sectors could exist, each with their own ``dark forces'' and constituent particles. Dark matter may be accompanied by hereto unknown gauge bosons (``dark force'' carriers,) which can couple dark matter particles and ordinary particles with exceptionally weak couplings.
Modern colliders can be blind to such new forces, even though the mass  of the ``dark force'' carriers can be quite small.
This is because the cross-sections of  relevant processes for ordinary matter are so small that the ``dark force'' events are simply statistically insignificant and are discarded in high-energy experiments.

Dark sector light weakly-coupled particles that interact with ordinary matter have been proposed as explanations of astronomical anomalies \cite{Fay04, ArkFinDou09} as well as discrepancies between calculated and measured muon magnetic moment \cite{Fay07,Pos09}. Such interactions would be inevitably below the weak force scale, ergo, the dark sector has so far escaped detection.
There are several proposed inroads into the detection of weakly-coupled particles and their associated dark forces~ \cite{Essig:2013lka}. One such example is the dark photon \cite{Hol86} that is hypothesized to be a massive particle which couples to electromagnetic currents just like the photon does. In addition, dark Z bosons have been proposed \cite{DavLeeMar14} that couple to the weak neutral currents, (i.e., their interactions are parity violating.) In a sense dark photons are massive photons while dark Z bosons are light-versions of Z bosons. From here on we will refer to this type of dark force mediator particle as the light gauge-boson.

Motivated by such ``dark force''  ideas here we place constraints both on couplings and masses of dark force carriers (light gauge bosons) by  reinterpreting results of experiments on searches for permanent electric dipole moments (EDMs) of diamagnetic atoms. Specifically we focus  on dark forces generated by the P,T-odd interaction of electrons and nucleons through the exchange of a massive light gauge boson.  We will refer to the carrier as $\chi$. Effectively, the usually-employed contact interactions are replaced with Yukawa-like interactions.

Standard Model (SM) predicts the existence of intrinsic permanent electric dipole moments (EDM) in particles as varied as quarks, leptons, and baryons.
These SM predictions, however, are below the current levels of experimental accuracy.
As an example, in the SM framework, the electron EDM is estimated to be of the order of $10^{-41}~\text{e~cm}$ (see e.g., Ref~\cite{Fuk12}) while the most stringent experimental limit stands at  $d_e<8.7 \times 10^{-29} ~\rm{e~cm}~(90\%~\rm{C.L.})$ from the ThO molecular search \cite{ACME14}. Remarkably, however, there are many theoretical extensions to the SM  that predict  EDM values  comparable to the present experimental constraints.


Overall, the searches for atomic EDM can be classified into two major categories: EDM of paramagnetic atoms and molecules and diamagnetic atoms.
Paramagnetic atoms, such as Tl and Cs, have an unpaired valence electron and the atomic EDM in this category is attributed to the EDM of the unpaired electron.
 Diamagnetic atoms, on the other hand, are closed-shell atoms. In discussions of diamagnetic atomic EDMs, the EDM is usually associated with the intrinsic EDM of an unpaired nucleon  (Schiff moment or P,T-odd electron-nucleon interactions). The best limit on a diamagnetic atom so far is~\cite{GriSwaLof09}
 \begin{equation}
 d(^{199} \text{Hg}) < 3.1 \times 10^{-29} \, \mathrm{e~cm} \, (95\%~\mathrm{C.L.}) \label{Eq:EDMHg}
\end{equation}
While we will use the Hg EDM result for putting constraints on the light mediators, the formalism and derived analytical expressions are applicable to other diamagnetic systems, such as the atoms of current experimental interest: xenon~\cite{GemHeiKar10}, ytterbium~\cite{RomFor99,Nat05},  radon~\cite{TarRanBal14}, and radium~\cite{GueSciAhm07,HolAhmBai10}.

\section{Basic setup}
We start by reviewing the structure of contact interactions formed out of products of bi-linear forms. The entire set of ten unique semi-leptonic Lorentz-invariant products is tabulated in Ref.~\cite{Khr91}.
In this paper we focus on the most commonly used parity, time-violating tensor current term (see, e.g., \cite{Mar85,DzuFlaPor09})
 \begin{align}
W_{T_e T_n}= &\int d\mathbf{r} \varepsilon_{\chi \lambda \mu \nu}[\bar{\psi}_{e}i g'^T_e \sigma_{\chi\lambda} \psi_{e}](\mathbf{r}) [\bar{\psi}_{n} i g'^T_n \sigma^{\mu \nu} \psi_{n}](\mathbf{r})\label{Eq:PToddTensor} \, .
\end{align}
Here  $\psi_{e,n}$ are the electron/nucleon Dirac bi-spinors and $\bar{\psi}_{e,n}=\psi_{e,n}^\dagger \gamma_0$ are their adjoints, while $g'^T_{e,n}$ are coupling constants.
Further, $\sigma_{\mu \nu}=\frac{1}{2}(\gamma_\mu\gamma_\nu-\gamma_\nu\gamma_\mu)$, where $\gamma$'s are Dirac matrices and $\varepsilon_{ijkl}$ is the four-dimensional Levi-Civita tensor. From Eq.(\ref{Eq:PToddTensor}) one could easily  read off the interaction Hamiltonians acting in the electron space by removing averaging over $\mathbf{r}$ and ``rubbing off'' $\psi_{e,n}$ and $\psi_{e,n}^\dagger$,
 \begin{equation*}
 h_{T_e T_n}(\mathbf{r}_e)=  \varepsilon_{\chi \lambda \mu \nu}[\gamma_0 i g'^T_e \sigma_{\chi\lambda} ]_e [\bar{\psi}_{n}(\mathbf{r}_e) i g'^T_n \sigma^{\mu \nu} \psi_{n}(\mathbf{r}_e) ] \label{Eq:H-PToddTensor} \, ,
 \end{equation*}
 where the subscript $e$ emphasizes that the  operators inside $[ \ldots]_e$ act on the electron degrees of freedom.

 The interaction~(\ref{Eq:PToddTensor}) is of the contact nature, i.e.,  it is constructed in the limit of the infinite mass of the carrier. For the finite mass $m_\chi$ of the mediator $\chi$ the interaction  need to be modified by sandwiching the currents with the Yukawa-type interaction (see e.g., Ref.~\cite{ZeeQFTBook})
  \begin{equation}\label{Eq:Yukawa}
 V_\chi (\mathbf{r},\mathbf{r'})=  \frac{e^{-m_\chi c |\bm{r}-\bm{r'}| }}{4\pi |\bm{r}-\bm{r'}|}\,.
  \end{equation}
 The ``upgraded'' Eq.~(\ref{Eq:PToddTensor})  reads
\begin{align}
W_{T_e T_n}^\chi  = & (m_\chi c)^2 \int \int d\mathbf{r}  d\mathbf{r'}
\varepsilon_{\chi \lambda \mu \nu}\, [\bar{\psi}_{e}i g'^T_e \sigma_{\chi\lambda} \psi_{e}](\mathbf{r})  \times \nonumber \\
&  V_\chi (\mathbf{r},\mathbf{r'})  [\bar{\psi}_{n} i g'^T_n \sigma^{\mu \nu} \psi_{n}](\mathbf{r'})
\label{Eq:PToddTensorChi}
\end{align}
or
\begin{align}
h_{T_e T_n}^\chi(\mathbf{r}_e)  = & (m_\chi c)^2
\varepsilon_{\chi \lambda \mu \nu}[\gamma_0 i g'^T_e \sigma_{\chi\lambda} ]_e \times \label{Eq:H-PToddTensorChiVerI}   \\
&  \int   d\mathbf{r}_n
 V_\chi (\mathbf{r}_e,\mathbf{r}_n)  [\bar{\psi}_{n}(\mathbf{r}_n)   i g'^T_n \sigma^{\mu \nu} \psi_{n}(\mathbf{r}_n) ] \, . \nonumber
\end{align}
 It is easily verified that in the limit of large propagator mass, $m_\chi$, the Yukawa potential in the above equation becomes $\delta^3(\bm{r}_e-\bm{r}_n)/(m_\chi c)^2$  recovering Eq.(\ref{Eq:PToddTensor}).
 We add a superscript $\chi$ to the interaction, ($W_{T_e T_n} \to W_{T_e T_n}^\chi$) to distinguish between the contact and finite-range interactions.

The structure of the expression (\ref{Eq:H-PToddTensorChiVerI}) suggests that the nuclear property  $[\bar{\psi}_{n}(\mathbf{r}_n)   i g'^T_n \sigma^{\mu \nu} \psi_{n}(\mathbf{r}_n) ]$ is ``carried out'' beyond the nucleus by the Yukawa potential. Thus one anticipates the P,T-odd forces would ``leak out'' of the nucleus on characteristic distances $$\lambda_\chi =1/(m_\chi c)$$ equal to the Compton wavelength of the mediator.

 We are interested in the atomic permanent electric dipole moments of diamagnetic systems induced by the P,T-odd semi-leptonic interaction.  The induced EDM $ \mathbf{d}$ of the atomic state $\Psi_0$  of energy $E_0$ can be expressed as
 \begin{equation}
 \mathbf{d}=\sum_{i}  \frac{\langle \Psi_0 |\bm{D}_e|\Psi_i \rangle \langle \Psi_i | H_{T_e T_n}^\chi |\Psi_0\rangle}{E_0-E_i} + \mathrm{c.c.} \, , \label{Eq:EDM-general}
\end{equation}
where c.c.\ stands for the complex conjugate of the preceding term and $E_i$ and $\Psi_i$ are the atomic energies and wave functions.
$H_{T_e T_n}^\chi= \sum_i h_{T_e T_n}^\chi (\mathbf{r}_i) $  and $\mathbf{D}_e = -|e| \sum_i \mathbf{r}_i$ is the operator of electric dipole moment for atomic electrons and the sum is over all atomic electrons.

 Tensor interaction~(\ref{Eq:H-PToddTensorChiVerI})  can be simplified further~\cite{Khr91} as  the nucleon motion can be well approximated as being non-relativistic.
 The result is
 \begin{align}
h_{T_e T_n}^\chi(\mathbf{r}_e)  = & - (m_\chi c)^2 4 i
[ i g'^T_e \gamma_0  \gamma_5 {\bm \sigma}]_e \cdot \label{Eq:H-PToddTensorChi}   \\
  \sum_{\rm nucleons} \int   d\mathbf{r}_n &
  V_\chi (\mathbf{r}_e,\mathbf{r}_n)     i g'^T_n \bm{ \sigma}_n \psi^\dagger_{n}(\mathbf{r}_n)  \psi_{n}(\mathbf{r}_n)  \, , \nonumber
\end{align}
i.e., it is proportional to the linear combination of weighted scalar products between nucleon and electron spins. Here we explicitly introduced  the summation over the nucleons. We further define
  \begin{align}
 &  \sum_{\rm nucleons} \int   d\mathbf{r}_n
  V_\chi (\mathbf{r}_e,\mathbf{r}_n)     i g'^T_n  \bm{\sigma}_n \psi^\dagger_{n}(\mathbf{r}_n)  \psi_{n}(\mathbf{r}_n)  \equiv  \nonumber\\
 & i \tilde{g}'^T_n  \bm{\sigma}_N  \int   d\mathbf{r}_n
  V_\chi (\mathbf{r}_e,\mathbf{r}_n)    \rho(\mathbf{r}_n) \, ,
 \end{align}
 since  $\psi^\dagger_{n}(\mathbf{r}_n)  \psi_{n}(\mathbf{r}_n)$ is the contribution of an individual nucleon to the nuclear density $\rho(\mathbf{r}_n)$.  The two sides of this equation can be related from nuclear structure calculations (see, e.g., Ref.~\cite{DzuFlaSil85}) which would define the constant $\tilde{g}'^T_n$.  Thereby,
the effective form of the interaction can be represented as a scalar product of the nuclear spin  and a rank-1 irreducible tensor acting in the electron space
 \begin{equation}
h_{T_e T_n}^\chi(\mathbf{r}_e)  = \bm{\sigma}_N \cdot \mathbf{t}_{e}^\chi
 \end{equation}
 with
 \begin{align}
\mathbf{t}_{e}^\chi  =& 4 \pi \sqrt{2} G_F  \, (m_\chi c)^2  \, C_{TN}(m_\chi)  \times \nonumber \\
& (i \gamma_0 \gamma_5 \bm{\sigma})_e  \int   d\mathbf{r}_n
  V_\chi (\mathbf{r}_e,\mathbf{r}_n)    \rho(\mathbf{r}_n) \, , \label{Eq:t-Y}
 \end{align}
 where  we introduced the  parameterization
 $ \tilde{g}'^T_n  g'^T_e= \pi \sqrt{2} G_F C_{TN}^\chi$ in terms of the Fermi constant, $G_F$, and the mass-dependent coupling constant $C_{TN}^\chi$. In the contact approximation this parameterization recovers the conventional form of the semi-leptonic operator ($C_{TN}^c \equiv \lim_{m_\chi \to \infty} C_{TN}^\chi )$
 \begin{equation}
\mathbf{t}_{e}^c \equiv \lim_{m_\chi \to \infty }
 \mathbf{t}_{e}^\chi  =\sqrt{2} G_F C_{TN}^c  \times (i \gamma_0 \gamma_5 \bm{\sigma} )_e     \rho(\mathbf{r}_e) \, .  \label{Eq:t-c}
 \end{equation}
 The finite-mass operator (\ref{Eq:t-Y}) can be recast in the  form analogous to the above equation,
 \begin{equation}
\mathbf{t}_{e}^\chi  = \sqrt{2} G_F C_{TN}^\chi  \times (i \gamma_0 \gamma_5 \bm{\sigma} )_e     \rho_\chi(\mathbf{r}_e) \, , \label{Eq:t-Y-rho}
 \end{equation}
by introducing  the effective ``Yukawa-weighted'' nuclear density
\begin{equation}
\rho_\chi(r) \equiv 4 \pi (m_\chi c)^2  \int   d\mathbf{r}_n
  V_\chi (\mathbf{r},\mathbf{r}_n)    \rho(\mathbf{r}_n)  \label{Eq:rho-chi} \, .
\end{equation}

The essential difference between the infinite-mass~(\ref{Eq:t-c})  and the finite-mass~(\ref{Eq:t-Y-rho}) cases is the replacement of the nuclear density $\rho(r)$ with the effective nuclear density $\rho_\chi(r)$. This effective nuclear density has been introduced earlier in Ref.~\cite{BouPik83} in the context of atomic parity violation mediated by a light gauge boson. It also plays an important role in our analysis.
For a uniform nuclear distribution contained inside a sphere of radius $R$ (i.e.,
$\rho(r<R)\equiv \rho_0 = 3/(4 \pi R^3)$) the effective nuclear density can be evaluated analytically  (see Appendix),
\begin{align}
    \rho_\chi(r)&= \rho_0 \, \frac{ \lambda_\chi}{ r} \times  \label{Eq:EffectiveDensity} \\
& \begin{cases}
&  \frac{r}{\lambda_\chi}-e^{- \frac{R}{\lambda_\chi} }(1+\frac{R}{\lambda_\chi})\sinh(\frac{r}{\lambda_\chi} ) ,\,  r\leq R,\\[2ex]
&  e^{-\frac{r}{\lambda_\chi}}\left(\frac{R}{\lambda_\chi} \cosh(\frac{R}{\lambda_\chi})-\sinh(\frac{R}{\lambda_\chi}) \right),\,  r> R. \nonumber
\end{cases}
\end{align}

\begin{figure}[ht]
\begin{center}
\includegraphics[width=0.5\textwidth]{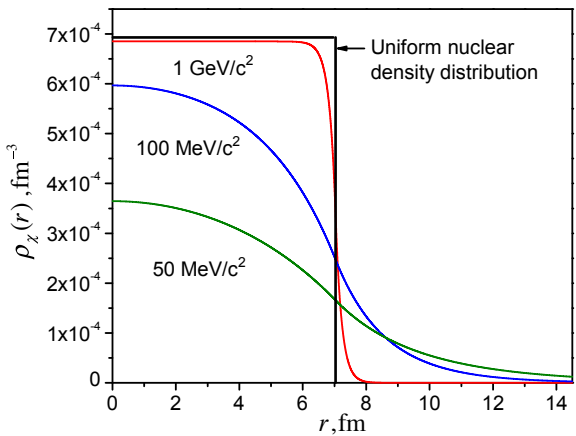}
\end{center}
\caption[Effective Nuclear Density Distribution for Hg]
  {$\rho_\chi(r)$, the effective ``Yukawa-weighted'' nuclear density for the $^{199}\textrm{Hg}$  nucleus is shown as a function of the radial distance $r$ for specified  values of $m_\chi$, the mass of the carrier. The nuclear cut-off radius is $R=7.0369$ fm.
  \label{Fig:MercDens}}
\end{figure}

Notice that outside  the nucleus, $\rho_\chi(r) \propto 1/r \, e^{-r/\lambda_\chi}$, i.e., as expected, the interaction~(\ref{Eq:t-Y}) would sample electronic cloud at distances $\lambda_\chi =\hbar/(m_\chi c)$ beyond the nuclear edge.
The values of $\lambda_\chi$ are (4,2,0.2) fm for  $m_\chi$ = (50,100,1000) MeV/c$^2$.
In Fig.~(\ref{Fig:MercDens}) we plot $\rho_\chi(r)$ for these mediator masses for the $^{199}\textrm{Hg}$ nucleus. The tendency of the effective density $\rho_\chi(r)$ to further ``leak out'' of the nucleus as the $m_\chi$ values are decreased is apparent.

When the range of the force,   $\lambda_\chi$, is comparable to an atomic size ($a_B$),  the interaction would extend over the entire atom and would sample atomic shell structure. This happens at the characteristic value of $m_\chi = \alpha m_e = 3.7 \,\mathrm{keV}/c^2$.

\section{Atomic structure}
Now we focus on the atomic-structure aspect of the problem.
$^{199}\textrm{Hg}$ is an 80-electron closed-shell system, with the electron configuration [Xe] $4f^{14} 5d^{10} 6s^2$.
One needs to evaluate the induced atomic EDM (\ref{Eq:EDM-general}) with the finite-mass mediator interaction (we refer to its value as $d_\chi$) and compare it with the contact-interaction result $d_c$.
In particular we focus on  the ratio
\begin{equation}
\mathcal{R}(m_\chi) \equiv \frac{d_\chi/C_{TN}^\chi }{d_c/C_{TN}^c} \label{Eq:Ratio} \, .
\end{equation}
We  employ two atomic-structure methods to evaluate this ratio: Dirac-Hartree-Fock (DHF) and Relativistic Random-Phase Approximation (RRPA).  RRPA improves upon DHF's independent-particle approximation by including major correlation effects. Both methods are {\em ab initio} relativistic, as they are based on solutions of the Dirac equation. The relativistic approach is important especially for large carrier masses for which the interaction is lumped in the nuclear region where the atomic electrons move at relativistic velocities. While there are more advanced techniques available~\cite{SinSah14}, the DHF and RRPA methods should provide an adequate qualitative understanding of how the atomic EDM responds to the finite-range forces.

In the independent-particle approximation (DHF), the induced atomic EDM (\ref{Eq:EDM-general})  becomes
\begin{equation}
\bm{d}_\chi=\bm{\sigma}_N \frac{2}{3} \sum_{as}(-1)^{j_a-j_s}\frac{\langle a||\bm{r}||s\rangle
\langle s|| \mathbf{t}_{e}^\chi  ||a\rangle}{\varepsilon_s-\varepsilon_a} \, ,
\label{Eq:EDM-angular}
\end{equation}
where the summation is carried over  atomic orbitals $a$ and $s$. $a$ are core orbitals
occupied in the ground state $\Psi_0$ and $s$ are un-occupied (excited or virtual) orbitals. $\varepsilon_i$ are the DHF energies of these orbitals. Each orbital $ \phi_{n\kappa m}$ is characterized by the principle quantum number $n$, relativistic angular momentum number $\kappa$ and magnetic quantum number $m$. $\kappa$  encodes the total angular momentum $j$ and the orbital angular momentum $\ell$.

The reduced matrix elements of  $ \mathbf{t}_{e}^\chi $ are
%
\begin{align}\label{Eq:t-Y-reduced}
&\langle n_a,\kappa_a||  \mathbf{t}_{e}^\chi  ||n_b, \kappa_b\rangle= -  \sqrt{2} G_F C_{TN}^\chi
\int^\infty_0 dr \rho_\chi(r) \times \\
& \left(\langle\kappa_a||\bm{\sigma}_e||-\kappa_b\rangle P_a(r) Q_b(r)+\langle-\kappa_a||\bm{\sigma}_e||\kappa_b\rangle Q_a(r) P_b(r) \right) \nonumber \,.
\end{align}
Here $P_{n\kappa}(r)$ and $Q_{n\kappa}(r)$ are the radial large and small components from the parameterization
\begin{align}
\label{Eq:WaveFunction}
 \phi_{n\kappa m}(\mathbf{r})=\frac{1}{r}\left(
                                                        \begin{array}{c}
                                                          i~P_{n\kappa}(r)\,\Omega_{\kappa m}(\hat{r})\\
                                                           Q_{n\kappa}(r)\,\Omega_{-\kappa m}(\hat{r}) \\
                                                        \end{array}
                                                      \right),
 \end{align}
with $\Omega_{\kappa m}$ being the spinor spherical harmonics.

Numerical procedure can be described as follows. First we solve the DHF equations for the ground state of Hg atom using the finite-differencing techniques~\cite{Joh07book}.
Next, we use the obtained DHF self-consistent potential to construct a finite basis set of atomic orbitals using the dual-kinetic-ballance $B$-spline technique~\cite{BelDer08}. This set of basis functions is finite and numerically complete (i.e., excited and continuum states are included in the set.) With such a set,  the  summation over atomic orbitals in Eq.(\ref{Eq:EDM-angular}) becomes a straightforward exercise.
In a typical calculation we use a set of basis functions expanded over 80 $B$-splines of order 9, in a cavity of spherical radius of 30 bohr and a 1000-point grid, out of which 68 points reside inside the nucleus, providing adequate numerical accuracy for both large and small carrier masses.

Compared to the DHF method, the more sophisticated  RRPA  approach accounts for a  linear response of an atom to a perturbing interaction ($h_{T_e T_n}^\chi(\mathbf{r}_e)$).  As a result of solving the RRPA equations~\cite{Joh88} using the described DHF basis set  we determined a quasi-complete set of particle-hole excited states and their energies, required for evaluating the sum over intermediate states in the EDM expression~(\ref{Eq:EDM-general}). The developed codes are an extension of the DHF and RRPA codes of Ref.~\cite{RavDer04}.

\section{Results}
\begin{figure}[ht]
\begin{center}
\includegraphics[width=0.5\textwidth]{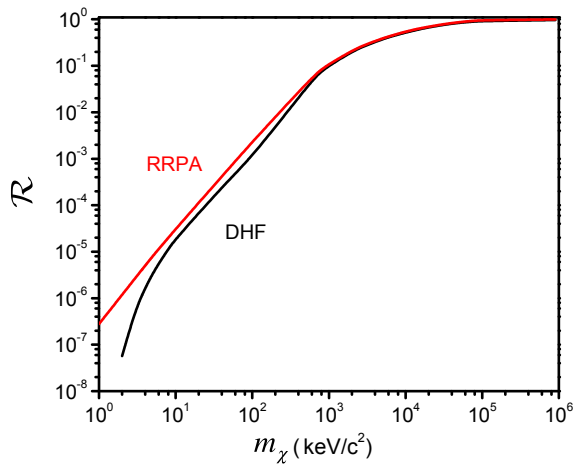}
\end{center}
\caption[EDM Ratios]
  {
   The dependence of the EDM ratio $\mathcal{R}$  for $^{199}$Hg atom on the carrier mass $m_\chi$.
  }\label{Fig:ratio}
\end{figure}

We start by describing numerical results for the ratio, $\mathcal{R}(m_\chi)$ (\ref{Eq:Ratio}), and then present analytical formulae. Our calculated ratios $\mathcal{R}(m_\chi)$ are plotted   in Fig.~\ref{Fig:ratio} as a function of the carrier mass $m_\chi$.


\subsection{Numerical approach}

To test the quality of the developed code, we first perform EDM calculations with the usual contact interaction of Eq.(\ref{Eq:t-c}), along with the nuclear Fermi distribution, in the DHF approximation.
 The resulting EDM, $\bm{d}(^{199} \text{Hg})=4.5 \times 10^{-12}\, C_{TN}^c  \bm{\sigma_N} \, \rm{a.u.}$, recovers the earlier result~\cite{DzuFlaPor09} obtained in the same approximation. Next, to simplify
 the integration in Eq.~(\ref{Eq:t-Y-reduced}), we replace the Fermi-type distribution with the uniform nuclear distribution, shown with the solid line in Fig.~\ref{Fig:MercDens}. Such calculation yields $\bm{d}(^{199} \text{Hg})=4.8 \times 10^{-12}\, C_{TN}^c  \bm{\sigma_N} \, \rm{a.u.}$ slightly deviating from the quoted Fermi-distribution value. Since all our calculations of the ratio (\ref{Eq:Ratio}) are carried out with the uniform nuclear density distribution, we fix this value as the infinite-carrier-mass value $d_c$. This result is consistent with the earlier value~\cite{DzuFlaPor09}

Next we perform the EDM calculations in the RRPA approximation. Our calculation with the uniform nuclear density distribution yields $d_c=11.2\times 10^{-12} C_{TN}^c  \bm{\sigma_N} \, {\rm a.u.}=5.9\times 10^{-20} C_{TN}^c  \bm{\sigma_N} \, \rm{e\cdot cm}$, agreeing with Ref.~\cite{DzuFlaPor09} value.


The results of our numerical calculation of the ratio $\mathcal{R}(m_\chi)$ are plotted  in Fig.~\ref{Fig:ratio} as a function of the carrier mass $m_\chi$.
The ratio tends to zero for small masses. It monotonically increases to unity as the mass increases, as in this limit the effective ``Yukawa-weighted'' nuclear density $\rho_\chi(r)$ approaches the true nuclear density $\rho(r)$ (see Fig.~\ref{Fig:MercDens}), thereby $d_\chi \to d_c$ and $\mathcal{R} \to 1$.
For small masses we clearly observe a constant slope on the log-log plot. We will comment on this scaling law below.

In general, the RRPA and DHF results are in a good agreement for large carrier masses (about 4\% agreement for $m_\chi >  1 \, \mathrm{MeV}/c^2$).  
The difference between two approaches starts to grow larger as with decreasing $m_\chi$ the force starts to probe the atomic shell structure bringing sensitivity to the details of treating electron-electron correlations. The most drastic difference arises at $m_\chi =1 \, \mathrm{keV}/c^2$ when the DHF ratio becomes negative while the RRPA ratio remains positive.

\subsection{Analytical approach}
\subsubsection{ Region $\lambda_\chi  \ll a_B/Z$ ($m_\chi\gtrsim\mathrm{MeV}$)}
\label{Sec:AnalyticalAboveMeV}
We find that for sufficiently large masses, the entire dependence of the ratio  $\mathcal{R}$  on  the carrier mass can be well approximated by taking only a single channel contribution in the EDM sum over states~(\ref{Eq:EDM-angular}). This is the contribution from the excitation of the outer-most occupied orbital $|a\rangle=6s_{1/2}$ to the excited orbitals $|s\rangle=np_{1/2}, (n=6,7,\ldots)$. The $6s_{1/2}$ orbital is the least bound leading to the smallest energy denominators. Moreover, as $j$ increases the electrons tend to reside less in the nuclear region due to increased centrifugal barrier thereby suppressing T,P-odd matrix elements.
This single-channel approximation is fully supported  by our numerical experimentation~\cite{HemanDissertation2014}.

The above observation motivates an analytical approach which consists in evaluating matrix elements of the P,T-odd interactions analytically. We use the fact that the matrix elements  are mostly accumulated in the region close to the nucleus. In this region, the large and small radial components of atomic orbitals~(\ref{Eq:WaveFunction}) can be approximated as
\begin{align}
\label{Eq:ShortDistanceApprox2}
P_{n\kappa}(r)&=\frac{\kappa~ \zeta}{|\kappa|}\frac{(\gamma-\kappa)}{\Gamma(2\gamma+1)}\sqrt{\frac{1}{Z\nu^3}}(2Zr)^{\gamma}, \\
Q_{n\kappa}(r)&=\frac{\kappa~ \zeta}{|\kappa|}\sqrt{\frac{1}{Z\nu^3}}\frac{Z}{c\Gamma(2\gamma+1)}(2Zr)^{\gamma} \, , \nonumber
\end{align}
where $\nu$ is the effective principal quantum number, $\nu=n-\sigma_\ell$, with $\sigma_\ell$ being the quantum defect.  $\zeta$ is the effective screened charge felt by the electron, e.g., for the valence orbital $\zeta=1$. $Z$ is the nuclear charge and $\gamma=\sqrt{\kappa^2- (\alpha Z)^2}$.
These formulae were adopted from Ref.~\cite{Khr91} for our parameterization~(\ref{Eq:WaveFunction}) of atomic orbital bi-spinors. Notice that these expressions were obtained for a point-like nucleus and they are valid for radial distances  $r \ll a_B/Z$ where the nuclear charge can be considered unscreened.

Now the reduced matrix element~(\ref{Eq:t-Y-reduced}) can be evaluated with the effective nuclear density~(\ref{Eq:EffectiveDensity}). While forming the ratio~(\ref{Eq:Ratio}) and limiting the summation to the single channel, we factor out the integrals
\begin{equation}
I(\rho_\chi) = \int_0^\infty  r^\beta \rho_\chi(r) dr \, ,
\label{Eq:Integral}
\end{equation}
which depend on the nuclear density. Such integrals do not depend on principal  quantum numbers, energies, nor dipole matrix elements and the ratio  can be simplified to
\begin{equation}
 \mathcal{R} = \frac{I(\rho_\chi)}{I(\rho)}
\end{equation}
with
 $\beta=2\sqrt{1- (\alpha Z)^2}$, because $|\kappa|=1$ for both the $s_{1/2}$ and $p_{1/2}$ orbitals.
Notice that this ratio does not depend on specific quantum numbers and it is valid as long as one of the excitation channels from an occupied  orbital $a$ ($n_a s_{1/2} \to p_{1/2}$ or $n_a p_{1/2} \to s_{1/2}$) is dominant. This argument is applicable to all diamagnetic atoms of current experimental interest: Xe, Yb, Hg, Rn, and Ra.

For the uniform nuclear density distribution we find the following formula ($u\equiv R/\lambda_\chi$)
\begin{align}
 \mathcal{R} &= 1+ (1+\beta) E_{1-\beta}(u) \left( \cosh(u) - \frac{1}{u} \sinh(u) \right) + \nonumber \\
 &-(1+u) e^{-u} \,_1\!F_2\left(   \frac{1}{2} + \frac{\beta}{2}; \frac{3}{2}, \frac{3}{2}+ \frac{\beta}{2}, \frac{ u^2}{4} \right) \label{Eq:analyticRatio} \, .
\end{align}
It is expressed in terms of  the generalized hypergeometric function $_1\!F_2$ and the exponential integral function $E_n(z) = \int_1^\infty e^{-zt}/t^n dt$.

For small values of the argument, i.e., in the limit $\lambda_\chi \gg R$  (yet $\lambda_\chi  \ll a_B/Z$)
\[
 \mathcal{R} \approx \frac{1}{3}  (1+\beta) \Gamma(\beta)  \left( \frac{R}{\lambda_\chi} \right)^{2-\beta}
 \xrightarrow[{\alpha Z \ll 1}]{}
 \left( \frac{R}{\lambda_\chi} \right)^{(\alpha Z)^2} \, ,
\]
where we also show the non-relativistic limit.
In the opposite case, $\lambda_\chi \ll R$,
\[
 \mathcal{R} \approx 1 - (1+\beta) (2-\beta ) \left( \frac{\lambda_\chi}{R} \right)^2
  \xrightarrow[{\alpha Z \ll 1}]{}1-3(\alpha Z)^2 \left( \frac{\lambda_\chi}{R} \right)^2
 \,,
\]
i.e., as expected, for large carrier masses the interaction becomes increasingly contact.
In this case the mass scaling of the ratio  from the above equation is
\[
 \mathcal{R}-1 \propto m_\chi^{-2} \, .
 \]

\begin{figure}[ht]
\begin{center}
\includegraphics[width=0.5\textwidth]{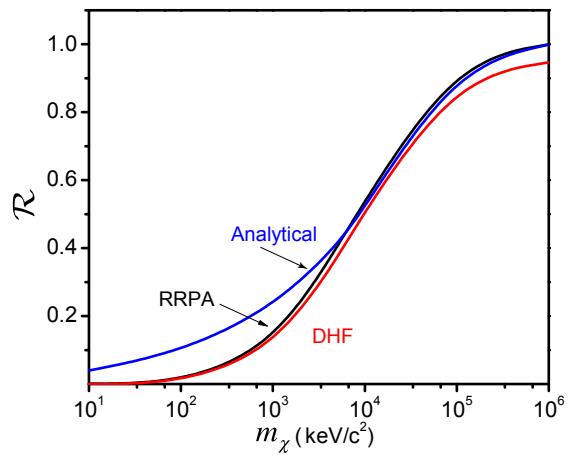}
\end{center}
\caption[EDM Ratios]
  { Comparison of numerical (DHF and RRPA) and analytical~(\ref{Eq:analyticRatio}) results for the Yukawa-to-contact-interaction EDM ratio $\mathcal{R}$  for $^{199}$Hg atom. The ratio is plotted as a function of the carrier mass.
  }\label{Fig:analyticalVsNumerical}
\end{figure}

We present the comparison between fully-numerical and analytical results in Fig.~\ref{Fig:analyticalVsNumerical} for mercury atom.
Now we would like to specify the region of validity of  the formula~(\ref{Eq:analyticRatio}).  First of all, the atomic wave functions~(\ref{Eq:WaveFunction}) were obtained using point-like nuclear charge distribution. In reality, the atomic orbitals are affected by the extended nuclear charge and inside the nucleus  the relevant product $P_{ns_{1/2}}Q_{n'p_{1/2}} \propto r^2$ instead of the ``softer'' dependence   $P_{ns_{1/2}}Q_{n'p_{1/2}} \propto r^\beta$ ($0<\beta<2$) used in the integral~(\ref{Eq:Integral}). Thus (\ref{Eq:analyticRatio}) would tend to de-emphasize the nuclear region.  Another limitation comes from the fact that the approximate wave functions are valid only in the region $r \ll a_B/Z$. This places constraints on the Compton wavelength of the force mediator $\lambda_\chi  \ll a_B/Z$, translating into $m_\chi \gg \alpha m_e/Z$ or $m_\chi \gg  0.3 \,\mathrm{MeV}/c^2$ for $^{199}$Hg, consistent with Fig.~\ref{Fig:ratio}.

\begin{figure}[ht]
\begin{center}
\includegraphics[width=0.45\textwidth]{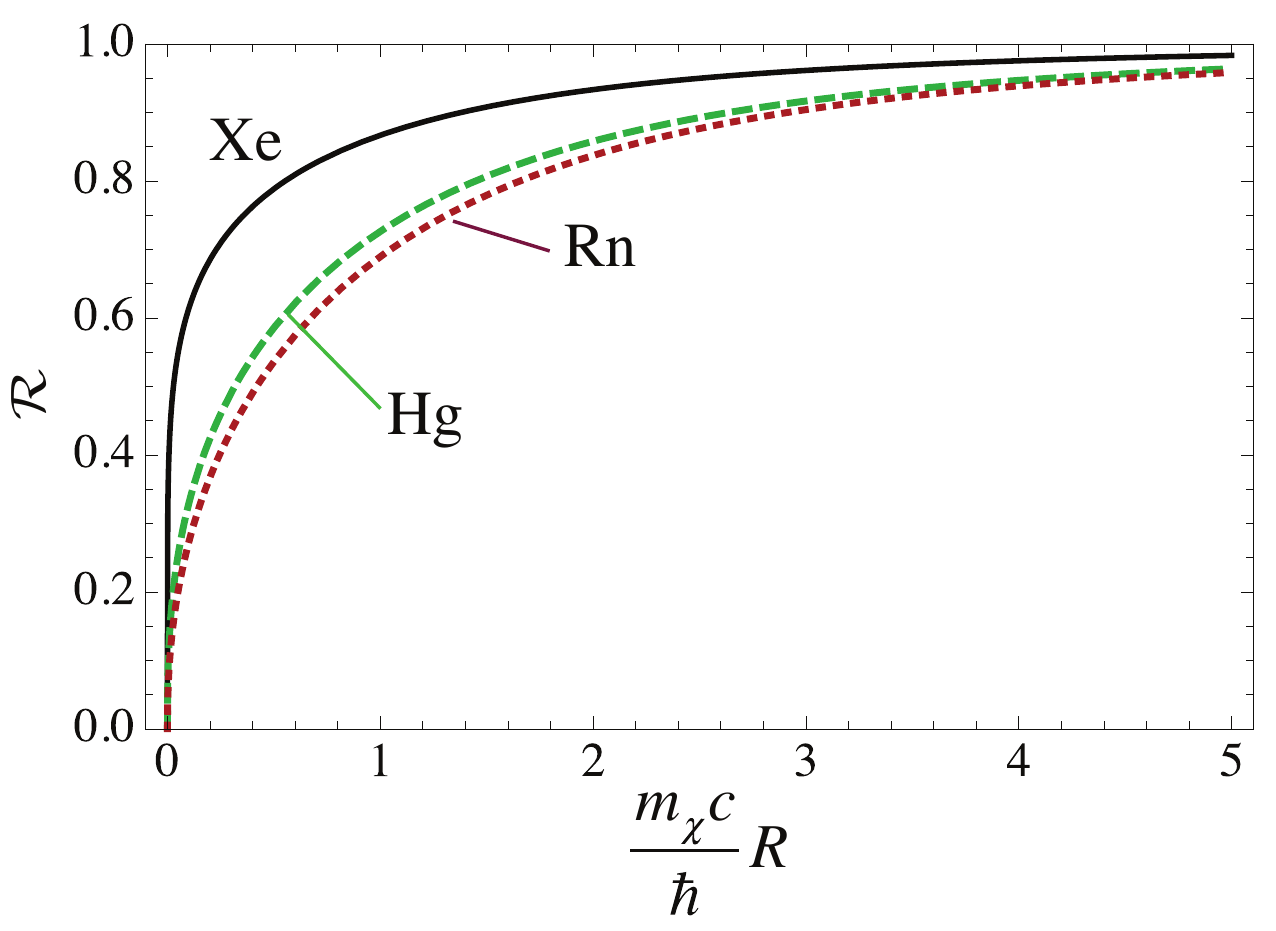}
\end{center}
\caption
  { (Color online) Yukawa-to-contact-interaction EDM ratios  $\mathcal{R}$, Eq.~(\ref{Eq:analyticRatio}) for xenon ($Z=54$, black solid  line), mercury ($Z=80$, green dashed line), and radon ($Z=86$, dotted red line) as a function of the ratio of the nuclear radius $R$ to the Compton wavelength of the force carrier. Notice that Eq.~(\ref{Eq:analyticRatio}) and the plotted curves  hold only for $R/\lambda_\chi \gg Z R/a_B \sim 4 \times 10^{-3}$.
 \label{Fig:ratio-Xe-Hg-Rn}}
\end{figure}

Fig.~\ref{Fig:ratio-Xe-Hg-Rn} compares ratios $\mathcal{R}$ computed with Eq.~(\ref{Eq:analyticRatio}) for xenon ($Z=54$), mercury ($Z=80$), and radon ($Z=86$). For smaller $Z$ the ratio tends to start off steeper at smaller masses $m_\chi$ and saturates earlier for larger values of $m_\chi$. It is worth emphasizing that Eq.~(\ref{Eq:analyticRatio}) and the  curves on Fig.~\ref{Fig:ratio-Xe-Hg-Rn}   hold only for $R/\lambda_\chi \gg Z R/a_B \sim 4 \times 10^{-3}$.

\subsubsection{ Region $\lambda_\chi \gg a_B$ (sub-$\mathrm{keV}$ carrier mass)}

To extend the  analytical treatment into the region of sub-keV masses, we notice that when the force range is much larger than the atomic size ($\lambda_\chi \gg a_B$), the Yukawa potential~(\ref{Eq:Yukawa}) becomes Coulomb-like since $e^{-|\bf{r}_e -\bf{r}_n|/\lambda_\chi} \approx 1$.
In this case the interaction is no longer resides near the nucleus, and the single-channel approximation introduced in Sec.~\ref{Sec:AnalyticalAboveMeV} may break down. However, we may still find the mass dependence of the ratio analytically. Indeed, for $\lambda_\chi \gg a_B$,
the effective nuclear distribution is  simply (see Appendix)
\begin{align}
    \rho_\chi(r)&= \rho_0  \left( \frac{R}{\lambda_\chi}\right)^2  \times
& \begin{cases}
& \frac{1}{2} -   \frac{1}{6}  \left( \frac{r}{R}\right)^2,\,  r\leq R,\\[2ex]
&  \frac{1}{3} \, \frac{ R}{ r} ,\,  r> R \,. \label{Eq:EffectiveDensityLimHugeLambda}
\end{cases}
\end{align}
Thereby we may simply factor out the entire mass dependence from the ratio and for very low masses, $m_\chi \ll \alpha m_e \approx 3.7 \, \mathrm{keV}/c^2$,
\begin{equation}
\mathcal{R} = A \times   \left( \frac{R}{\lambda_\chi}\right)^2 \propto m_\chi^2 \, ,
\label{Eq:RveryLowMasses}
\end{equation}
where the mass-independent proportionality constant $A$ has to be evaluated with atomic-structure techniques,
$ A \equiv (\lambda_\chi/R)^2 \mathcal{R}  $,
with  the effective density (\ref{Eq:EffectiveDensityLimHugeLambda}).
For $^{199}$Hg we find $A=-49$ in the DHF approximation and {\bf $A=215$} in the more accurate RRPA method.

To summarize this section, the entire dependence of $\mathcal{R}$ on $R/\lambda_\chi$ can be described analytically with Eq.~(\ref{Eq:analyticRatio}) for masses above $\sim$  MeV and with Eq.~(\ref{Eq:RveryLowMasses}) for $m_\chi$ below $\sim$ keV. The values of $\mathcal{R}$  in the transition region between  these two limits depend on the atomic-shell structure of specific atom.

\section{Conclusions}

\begin{figure}[ht]
\begin{center}
\includegraphics[width=0.5\textwidth]{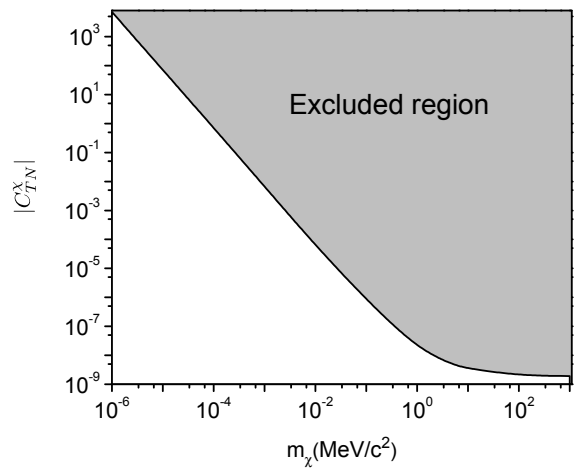}
\end{center}
\caption
  {  Exclusion region parameterized in terms of coupling strengths $|C_{TN}^\chi|$  and carrier masses $m_\chi$. This exclusion region is  derived from computed Yukawa-to-contact-interaction EDM ratio $\mathcal{R}$ in the RRPA method and the experimental limit~\cite{GriSwaLof09} on EDM of $^{199}\mathrm{Hg}$ atom. \label{Fig:limits}}
\end{figure}

Now with the computed Yukawa-to-contact-interaction EDM ratio $\mathcal{R}$, Eq.~(\ref{Eq:Ratio}), we proceed to placing constraints on the coupling strengths and masses of the light gauge bosons. Essentially, we require
\[
d_\chi = C_{TN}^\chi  \frac{d_c}{C_{TN}^c } \mathcal{R} < \text{Experimental limit on atomic EDM.}
\]
To place the limits on the coupling constant $C_{TN}^\chi$ we use the ratio $\mathcal{R}$  computed in the more sophisticated RRPA approach together with the the atomic EDM for contact interaction taken from Ref.\cite{DzuFlaPor09}. 
The experimental limit~\cite{GriSwaLof09} on Hg atom EDM reads
$|d(^{199}\mathrm{Hg})|<3.1 \times 10^{-29} \, e \, \mathrm{cm}$ (95\% C.L.). Thereby,
\[
|C_{TN}^\chi|  < |d(^{199}\mathrm{Hg})|  \left| \frac{C_{TN}^c }{d_c} \frac{1}{\mathcal{R}}  \right| \,.
\]
The resulting exclusion region for  $1 \, \mathrm{eV} <m_\chi < 1\, \mathrm{GeV}$  is shown in Fig.~\ref{Fig:limits}. The exclusion region can be trivially extended to the lower masses using Eq.~(\ref{Eq:RveryLowMasses}) (basically continuing the straight line on the log-log plot, Fig.~\ref{Fig:limits}). For higher masses $|C_{TN}^\chi|$ saturates to $|C_{TN}^c|\leq 1.9 \times 10^{-9}$.
The bounds on $|C_{TN}^\chi|$ become less stringent for lighter carriers due to the fact that
 as the range of the interaction becomes larger than the atomic size, the effective nuclear density scales down as $m_\chi^2$, Eq.~(\ref{Eq:EffectiveDensityLimHugeLambda}), reducing the atomic EDM enhancement factor.  For a fixed experimental limit on EDM, this translates into larger values of the  coupling constant $|C_{TN}^\chi|$.

\appendix
\section{The effective ``Yukawa-weighted'' nuclear density}
The effective ``Yukawa-weighted'' nuclear density is defined as
\begin{equation}\label{ApEq:YukawaND1}
  \rho_\chi(r)\equiv \frac{4\pi}{\lambda_\chi^2} \int d \bm{r}_n  V_\chi (\bm{r}, \bm{r}_n)\rho(\bm{r}_n),
\end{equation}
where the Yukawa potential $V_\chi(\bm{r}, \bm{r}_n)$ can be expanded as
\begin{align}\label{ApEq:Yukawa_potentioal}
  V_\chi(\bm{r}, \bm{r}_n)&=\frac{e^{-|\bm{r}-\bm{r}_n|/\lambda_\chi}}{4 \pi|\bm{r}-\bm{r}_n|} =\notag \\
 &- \lambda_\chi^{-1} \sum_{l=0}^{\infty} j_l(i{r}_</\lambda_\chi)h^{(1)}_l(i r_>/\lambda_\chi)\notag\\ & \times \sum_{m=-l}^lY^*_{lm}(\hat{r})Y_{lm}(\hat{r}_n),
\end{align}
where  $r_> = \max(r,r_n)$, $r_< = \min(r,r_n)$.
For a spherically-symmetric nuclear distribution the angular part of the integral in Eq.(\ref{ApEq:YukawaND1}) is reduced to
\begin{equation}
  \int Y^*_{lm}(\hat{r})Y_{lm}(\hat{r}_n) d \Omega_n=\frac{1}{4 \pi} \delta_{l0} \delta_{m 0} \, ,
\end{equation}
i.e., only the monopole contribution remains in Eq.(\ref{ApEq:Yukawa_potentioal}).
The Bessel and Hankel functions of imaginary arguments are the modified Bessel functions:
\begin{align}
  j_0(iz)&=i_0(z)=\frac{\sinh z}{z}, \\
  h^{(1)}_0(iz)&=-k_0(z)=-\frac{e^{-z}}{z} \,.
\end{align}
Therefore, we can rewrite Eq.(\ref{ApEq:YukawaND1}) as
\begin{equation}\label{ApEq:YukawaND2}
  \rho_\chi(r)= \frac{\rho_0}{\lambda_\chi^3} \int_0^\infty i_0(r_</\lambda_\chi) k_0(r_>/\lambda_\chi)r_n^2 d r_n \,.
\end{equation}
For a uniform nuclear distribution contained inside a sphere of radius $R$, (i.e.,
$\rho(r<R)\equiv \rho_0 = 3/(4 \pi R^3)$) the integral yields
\begin{align}
    \rho_\chi(r)&= \rho_0 \, \frac{ \lambda_\chi}{ r} \times  \label{Eq:EffectiveDensity} \\
& \begin{cases}
&  \frac{r}{\lambda_\chi}-e^{- \frac{R}{\lambda_\chi} }(1+\frac{R}{\lambda_\chi})\sinh(\frac{r}{\lambda_\chi} ) ,\,  r\leq R,\\[2ex]
&  e^{-\frac{r}{\lambda_\chi}}\left(\frac{R}{\lambda_\chi} \cosh(\frac{R}{\lambda_\chi})-\sinh(\frac{R}{\lambda_\chi}) \right),\,  r> R. \nonumber
\end{cases}
\end{align}
Notice that in the limit $\lambda_\chi \gg R$ (i.e., for $m_\chi \ll 10 \, \mathrm{keV}/c^2$),
\begin{align}
    \rho_\chi(r)&= \rho_0 \times
& \begin{cases}
& \frac{1}{2}  \left( \frac{R}{\lambda_\chi}\right)^2 -   \frac{1}{6}  \left( \frac{r}{\lambda_\chi}\right)^2,\,  r\leq R,\\[2ex]
&  \frac{1}{3} \left( \frac{R}{\lambda_\chi}\right)^3\, \frac{ \lambda_\chi}{ r} \,e^{-\frac{r}{\lambda_\chi}}  ,\,  r> R \,. \label{Eq:EffectiveDensityLimLargeLambda}
\end{cases}
\end{align}
Finally, in the limit $\lambda_\chi \gg a_B$ (i.e., the range of the potential being much larger than the atomic size), we could drop the exponent in the second line of Eq.~(\ref{Eq:EffectiveDensityLimLargeLambda})
\begin{align}
    \rho_\chi(r)&= \rho_0  \left( \frac{R}{\lambda_\chi}\right)^2  \times
& \begin{cases}
& \frac{1}{2} -   \frac{1}{6}  \left( \frac{r}{R}\right)^2,\,  r\leq R,\\[2ex]
&  \frac{1}{3} \, \frac{ R}{ r} ,\,  r> R \,. \label{Eq:App:EffectiveDensityLimHugeLambda}
\end{cases}
\end{align}

\acknowledgments
We would like to thank M. Pospelov for bringing this problem to our attention and his helpful comments on the manuscript. The work  was supported in part by the US National Science Foundation.

%

\end{document}